\documentclass[a4paper]{article}

\usepackage{INTERSPEECH2020}
\usepackage{cite}
\usepackage{url}

\DeclareMathOperator{\sigmoid}{sigmoid}
\DeclareMathOperator{\relu}{ReLU}
\DeclareMathOperator{\linear}{Linear}

\DeclareMathOperator{\ce}{CE}

\title{A Transformer-based Audio Captioning Model with Keyword Estimation}
\name{Yuma Koizumi, Ryo Masumura, Kyosuke Nishida, Masahiro Yasuda, and Shoichiro Saito}
\address{NTT Corporation, Japan}
\email{koizumi.yuma@ieee.org}

\begin{document}

\maketitle
\begin{abstract}
One of the problems with automated audio captioning (AAC) is the indeterminacy in word selection corresponding to the audio event/scene. Since one acoustic event/scene can be described with several words, it results in a combinatorial explosion of possible captions and difficulty in training. To solve this problem, we propose a Transformer-based audio-captioning model with keyword estimation called \textit{TRACKE}. It simultaneously solves the word-selection indeterminacy problem with the main task of AAC while executing the sub-task of acoustic event detection/acoustic scene classification (i.e., keyword estimation). TRACKE estimates keywords, which comprise a word set corresponding to audio events/scenes in the input audio, and generates the caption while referring to the estimated keywords to reduce word-selection indeterminacy. Experimental results on a public AAC dataset indicate that TRACKE achieved state-of-the-art performance and successfully estimated both the caption and its keywords.
\end{abstract}

\section{Introduction}
\label{sec:intro}

Automated audio captioning (AAC) is an intermodal translation task when translating an input audio into its description using natural language \cite{ac1,ac2,ac3,audiocaps,clotho,ntt_task6}. In contrast to automatic speech recognition (ASR), which converts a speech to a text, AAC converts environmental sounds to a text. This task potentially raises the level of automatic understanding of sound environment from merely tagging events \cite{aed,aed2} (e.g.\,alarm), scenes \cite{asc} (e.g.\,kitchen) and condition \cite{asd} (e.g.\,normal/anomaly) to higher contextual information including concepts, physical properties, and high-level knowledge. For example, a smart speaker with an AAC system will be able to output ``{\it a digital alarm in the kitchen has gone off three times},'' and might give us more intelligent recommendations such as ``{\it turn the gas range off}.''

One of the problems with AAC is the existence of many possible captions that correspond to an input. In ASR, a set of phonemes in a speech corresponds almost one-to-one to a word. In contrast, one acoustic event/scene can be described with several words, such as \{{\it car, automobile, vehicle, wheels}\} and \{{\it road, roadway, intersection, street}\}. Such indeterminacy in word selection leads to a combinatorial explosion of possible answers, making it almost impossible to estimate the ground-truth and difficulty in training an AAC system.

To reduce the indeterminacy in word selection, conventional AAC setups allow the use of keywords related to acoustic events/scenes \cite{audiocaps,clotho}. The audio samples in the AudioCaps dataset \cite{audiocaps} are parts of the Audio Set \cite{audioset}, and their captions are annotated while referring to the Audio Set labels. Therefore, automatic text generation while referring to keywords (e.g.\,Audio Set label) may restrict the solution space and should be effective in reducing word-selection indeterminacy.

Unfortunately, in some real-world applications such as using a smart speaker, it is difficult to provide such keywords in advance. For example, to output the caption ``{\it a digital alarm in the kitchen has gone off three times},'' conventional AAC systems require the keywords related to the acoustic events/scenes such as \{alarm, kitchen\}. However, if the user can input such keywords, the user should know the sound environment without any captions. This dilemma means that we need to solve the word-selection indeterminacy problem of AAC while simultaneously executing the traditional sub-task of acoustic event detection (AED) \cite{aed,aed2}/acoustic scene classification (ASC) \cite{asc}\footnote{A conventional method \cite{audiocaps} uses ASC-aware acoustic features such as the bottleneck feature of VGGish \cite{vggish}. In contrast, we attempt to solve the word-selection indeterminacy problem of AAC explicitly by using the AED/ASC sub-task.}.

\begin{figure}[t]
  \centering
\includegraphics[width=80mm,clip]{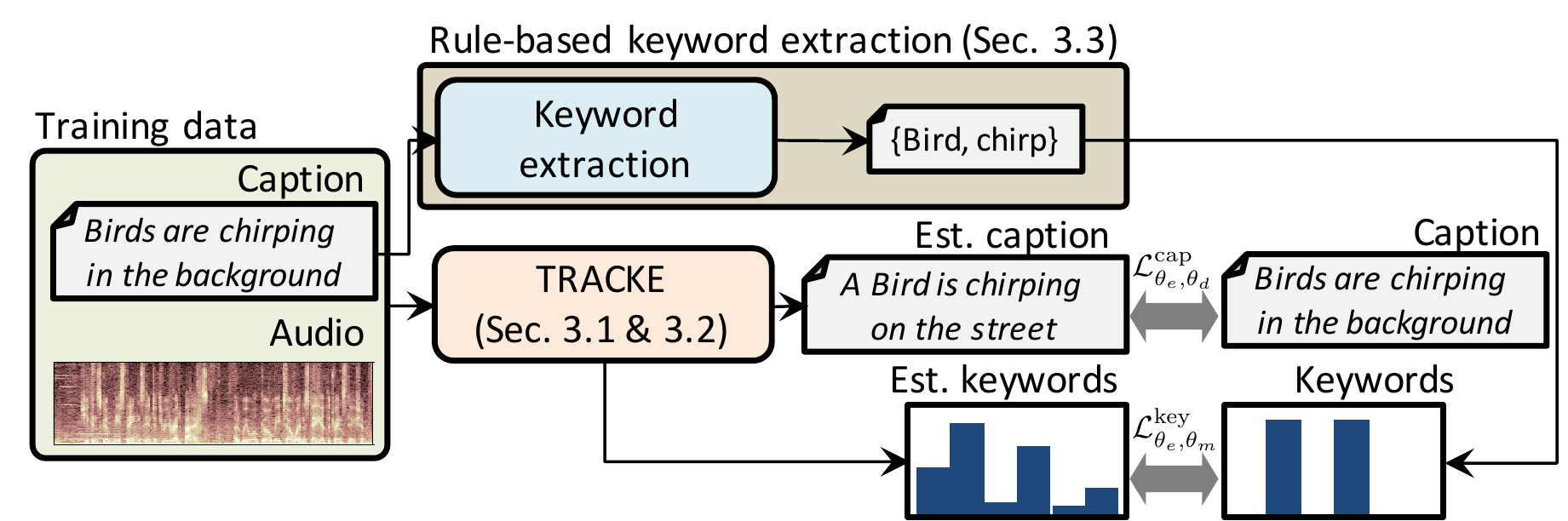} 
  \vspace{-15pt}
  \caption{Overview of training procedure of TRACKE.}
  \label{fig:ov}
  \vspace{-10pt}
\end{figure}

We propose a Transformer\cite{transformer}--based audio captioning model with keyword estimation called \textit{TRACKE}, which simultaneously solves the word-selection indeterminacy problem of AAC and executing the AED/ASC sub-task (i.e.\,keyword estimation). Figure\,\ref{fig:ov} shows an overview of the training procedure of TRACKE. TRACKE's encoder has a branch for keyword estimation and its decoder generates captions while referring to the estimated keywords for reducing word-selection indeterminacy. In the training phase, a set of ground-truth keywords is extracted from the ground-truth caption, and the branch is trained to minimize the estimation error of the keywords. A summary of our contributions is as follows.
\begin{enumerate}
 \setlength{\parskip}{0cm} 
 \setlength{\itemsep}{0.1cm} 
\item We decompose AAC into a combined task of caption generation and keyword estimation, and keyword estimation is executed by adopting a weakly supervised polyphonic AED strategy \cite{wpoly}.
\item This is the first study that has adopted Transformer \cite{transformer} to AAC\footnote{The use of a Transformer in AED/ASC tasks has been investigated \cite{transformer_aed,cnn_trans}.}. We also extended Transformer to simultaneously solve the word-selection indeterminacy problem of AAC and the related AED/ASC sub-task.
\end{enumerate}

\section{Preliminaries of audio captioning}
\label{sec:conv}

AAC is a task to translate an input audio sequence $( \bm{\phi}_1, ..., \bm{\phi}_T)$ into a word sequence $(w_1,...,w_N)$. Here, $\bm{\phi}_t \in \mathbb{R}^{D_x}$ is a set of acoustic features at time index $t$, and $T$ is the length of the input sequence. The output of AAC $w_n \in \mathbb{N}$ denotes the $n$-th word's index in the word vocabulary, and $N$ is the length of the output sequence.

Previous studies addressed AAC using a sequence-to-sequence model (seq2seq) \cite{seq2seq1,seq2seq2}. First, the encoder $\mathcal{E}$ embeds the input sequence into a feature-space as $\bm{\nu}$. Here, $\bm{\nu}$ can be either a fixed dimension vector or a hidden feature sequence. Then the decoder $\mathcal{D}$ predicts the posterior probability of the $n$-th word under the given input and 1st to $(n-1)$-th outputs recursively as
\begin{align}
\bm{\nu} 
&= \mathcal{E}_{\theta_e} \left( \bm{\phi}_1, ..., \bm{\phi}_T \right),
\label{eq:encoder}\\
p( w_n | \bm{\nu}, \bm{w}_{n-1} )
&= \mathcal{D}_{\theta_d} \left( \bm{\nu}, \bm{w}_{n-1} \right),
\label{eq:decoder}
\end{align}
where $\theta_e$ and $\theta_d$ are the sets of parameters of $\mathcal{E}$ and $\mathcal{D}$, respectively, $\bm{w}_{n-1} = (w_1,...,w_{n-1})$, and $w_n$ is estimated from the posterior using beam search decoding.

As mentioned above, one of the problems with AAC is indeterminacy in word selection. Since one acoustic event/scene can be described with several words, the number of possible captions becomes huge due to combinatorial explosion. To reduce such indeterminacy, previous studies used meta information such as keywords \cite{audiocaps,clotho}. We define $\bm{m} = \{ m_k \in \mathbb{N} \}_{k=1}^K$ as a set of keywords where $K$ is the number of keywords. By passing $\bm{m}$ to the decoder, it is expected that $\bm{m}$ works as an attention factor to select the keyword from the possible words corresponding to the acoustic event/scene. Thus, (\ref{eq:decoder}) can be rewritten as 
\begin{align}
p( w_n | \bm{\nu}, \bm{m}, \bm{w}_{n-1} )
= \mathcal{D}_{\theta_d} \left( \bm{\nu}, \bm{m}, \bm{w}_{n-1} \right).
\label{eq:posterior}
\end{align}

\section{Proposed Model}
\label{sec:prop}
In real-world applications, there are not many use-cases for AAC systems that require keywords. If the user can input such keywords, he/she should know the sound environment without any captions. To expand the use-cases of AAC, TRACKE generates a caption while estimating its keywords from the input audio. Sections \ref{sqe:prp:ov} and \ref{sec:prp:detail} give an overview and details of TRACKE, respectively, and Section \ref{sec:prp:key_gen} describes the procedure for extracting ground-truth keywords from the ground-truth caption.

\begin{figure}[t]
\centering
\includegraphics[width=85mm,clip]{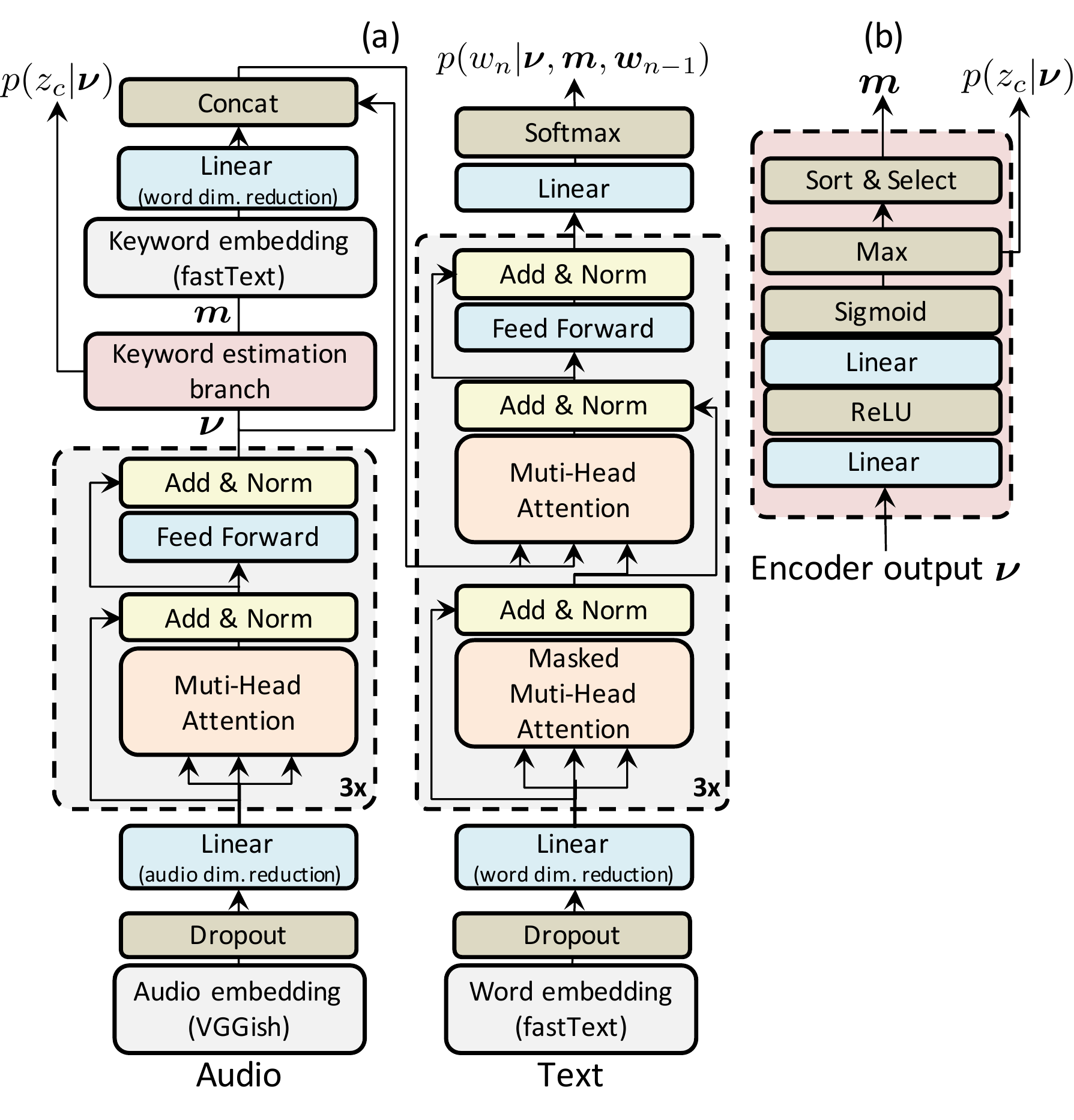} 
\vspace{-15pt}
\caption{(a) Architecture of TRACKE and (b) details of keyword-estimation branch $\mathcal{M}$.}
\label{fig:model}
\vspace{-15pt}
\end{figure}

\subsection{Model overview}
\label{sqe:prp:ov}

Figure\,\ref{fig:model} shows the architecture of TRACKE. The components of the encoder and decoder are the same of those of the original Transformer \cite{transformer}, but the number of stacks and hidden dimensions different. We use the bottleneck feature of VGGish \cite{vggish} ($D_x = 128$) for audio embedding, and fastText \cite{fastText} trained on the Common Crawl corpus ($D_w = 300$) for caption-word and keyword embedding, respectively. Since the dimension[s?] of audio feature and word embedding differ, we use two linear layers to adjust the dimensions of audio and word/keyword embedding to $D_f = 100$, which is the hidden dimension of the encoder/decoder.

In TRACKE, the size of the encoder output $\bm{\nu}$ is $D_f \times T$. The $\bm{\nu}$ is passed to the keyword-estimation branch $\mathcal{M}$ as
\begin{align}
\hat{\bm{m}} &= \mathcal{M}_{\theta_m} \left( \bm{\nu} \right),
\end{align}
where $\hat{\bm{m}} = \{ \hat{m}_k \in \mathbb{N} \}_{k=1}^K$ is the set of the estimated keywords, and $\theta_m$ is the parameter of $\mathcal{M}$. First, to input $\hat{\bm{m}}$ to $\mathcal{D}$, $\hat{\bm{m}}$ is embedded into the feature space using fastText word embedding. To adjust the feature dimension, the embedded keywords are then passed to the linear layer for dimension reduction of words/keywords. Then, the output $\mathbb{R}^{D_f \times K}$ is concatenated to $\bm{\nu}$. Finally, the concatenated feature $\mathbb{R}^{D_f \times (T+K)}$ is used as the key and value of the multi-head attention layers in $\mathcal{D}$, and the decoder estimates the posterior of the $n$-th word, the same as in (\ref{eq:posterior}), as
\begin{align}
p( w_n | \bm{\nu}, \hat{\bm{m}}, \bm{w}_{n-1} )
= \mathcal{D}_{\theta_d} \left( \bm{\nu}, \hat{\bm{m}} , \bm{w}_{n-1} \right).
\end{align}

\subsection{Keyword-estimation branch}
\label{sec:prp:detail}

Let $C$ be the size of the keyword vocabulary and $\bm{m}$ be a set of keywords extracted from the ground-truth caption (described in Section \ref{sec:prp:key_gen}). The $\mathcal{M}$ estimates $\bm{m}$, that is, whether the input audio includes audio events/scenes corresponding to keywords in the keyword vocabulary.

The duration of each event/scene is different, e.g., a passing train sound is long, while a dog barking is short. Thus, as in polyphonic AED \cite{poly1,poly2}, it would be better to estimate whether the pre-defined $c$-th event/scene has happened for each $t$. However, the given keyword labels are weak; start and stop time indexes are not given. Therefore, we carry out keyword estimation through the weakly supervised polyphonic AED strategy \cite{wpoly} by (i) estimating the posterior of each event on each $t$, $p( z_{c,t} | \bm{ \nu } )$, then (ii) aggregating these posteriors for all $t$, $p( z_c | \bm{ \psi } ) $. Then, the most likely $K$ events/scenes (i.e.\,keywords) are selected.

First is the posterior-estimation step; $\mathcal{M}$ estimates the posterior of the $c$-th keyword at $t$ as
\begin{align}
\hat{\bm{Z}} = 
\sigmoid
\left( \linear 
\left( \relu 
\left( 
\linear 
\left(
\bm{ \nu } 
\right)
\right) 
\right) 
\right) 
\label{eq:M_linears},
\end{align}
where $\hat{\bm{Z}} \in [0,1]^{C \times T}$ and its $(c,t)$ element is $p( z_{c,t} | \bm{ \nu } )$. Next is the posterior-aggregation step. We use the global max pooling strategy as follows because the maximum value rather than the average for considering the difference in the duration for each event
\begin{align}
p( z_c | \bm{ \nu } ) 
\approx 
\max _t \left[
p( z_{c,t} | \bm{ \nu } )
\right].
\label{eq:max_M}
\end{align}
Then, $p( z_c | \bm{ \nu } )$ is sorted in descending order and the top-$K$ keywords with high posterior are selected as $\hat{\bm{m}}$.

Note that the estimated order of the top-$K$ keywords has no effect on text generation because position encoding is not applied to the embedding vector of $\hat{\bm{m}}$. In addition, the computational graph is not connected from $\mathcal{M}$ to $\mathcal{D}$ because the sorting and top-$K$ selection after (\ref{eq:max_M}) are not differentiable. Therefore, text-generation loss is not back-propagated to two linear layers in (\ref{eq:M_linears}), i.e.\, the update of $\theta_m$ is only affected by the accuracy of keyword estimation.

\subsection{Rule-based keyword extraction for training}
\label{sec:prp:key_gen}

We describe a rule-based keyword extraction for generating $\bm{m}$. The keyword-estimation problem has been tackled as a sub-task of text summarization and comprehension, and several machine learning-based methods have been proposed \cite{text_rank,keyword01,keyword02,keyword03,keyword04}. 
In this study, the first attempt to reveal whether the use of estimated keywords is effective for AAC, we adopted a simple rule-based keyword extraction method.

We use frequent word lemmas of nouns, verbs, adjectives, and adverbs as keywords. From all captions in the training data, we first extract words that belong to the four parts of speech. Next, these words are converted to their lemmas and counted. Then, the keyword vocabulary is constructed using the most frequent $C$ lemmas except ``be''. Finally, the word lemmas that exist in the keyword vocabulary are used as the ground-truth keywords $\bm{m}$. In the case of a ground-truth caption in the Clotho dataset \cite{clotho} ``{\it A muddled noise of broken channel of the TV}'', the words that belong to the four target parts of speech are \{muddled, noise, broken, channel, TV\}. These words are then converted to their lemmas as \{muddle, noise, break, channel, TV\}. Finally, the lemmas that exist in the keyword vocabulary are extracted as $\bm{m}$.

\subsection{Training procedure}
\label{sec:prp:training}
TRACKE is trained to minimize two cost functions simultaneously; for captioning $\mathcal{L}^{ \mbox{\scriptsize cap} }_{\theta_e, \theta_d}$ and keyword estimation $\mathcal{L}^{ \mbox{\scriptsize key} }_{\theta_e, \theta_m}$. For $\mathcal{L}^{ \mbox{\scriptsize cap} }_{\theta_e, \theta_d}$, we used the basic cross-entropy loss as $\mathcal{L}^{\mbox{\scriptsize cap} }_{\theta_e, \theta_d} = N^{-1} \sum_{n=1}^N \ce \left( w_n, p( w_n | \bm{\nu}, \hat{\bm{m}}, \bm{w}_{n-1} ) \right),$ where $\ce$ is the cross-entropy between a given label and estimated posterior. For $\mathcal{L}^{ \mbox{\scriptsize key} }_{\theta_e, \theta_m}$, to avoid $\mathcal{M}$ from always outputting the most frequent keywords, we calculate weighted binary cross-entropy, the weight of which is the reciprocal of the prior probability, as 
\begin{align}
\mathcal{L}^{ \mbox{\scriptsize key} }_{\theta_e, \theta_m} = 
- \frac{1}{C} \sum_{c=1}^C 
\lambda_c z_c \ln \hat{z}_c
+
\gamma_c (1-z_c) \ln (1 - \hat{z}_c),
\end{align}
where $ \hat{z}_c = p( z_c | \bm{ \psi } ) $, and $z_c = 1$ if $c \in \bm{m}$; otherwise, $z_c = 0$. Here, $\lambda_c $ and $\gamma_c$ are the weights as $\lambda_c = ( p(z_c) )^{-1}$ and $\gamma_c = ( 1 - p(z_c) )^{-1}$, respectively, where $p(z_c)$ is the prior probability of the $c$-th keyword calculated by
\begin{align}
p(z_c) = \frac{\mbox{\# of $c$-th keyword in training captions}}{ \mbox{\# of training captions} }.
\end{align}

\section{Experiments}
\label{sec:exp}

\subsection{Experimental setup}

\textbf{Dataset and metrics:} We evaluated TRACKE on the Clotho dataset \cite{clotho}, which consists of audio clips from the Freesound platform \cite{freesound} and its captions were annotated via crowdsourcing \cite{crowd}. 
This dataset was used in a challenge task of the Detection and Classification of Acoustic Scenes and Events (DCASE) 2020 Challenge \cite{dcase2020}. 
We used the development split of 2893 audio clips with 14465 captions (i.e.\,one audio clip has five ground-truth captions) for training and the evaluation split of 1045 audio clips with 5225 captions for testing. 
From the development split, 100 audio clips and their captions were randomly selected as the validation split. 
We evaluated TRACKE and three other models on the same metrics used in the DCASE 2020 Challenge, i.e., BLEU-1, BLEU-2, BLEU-3, BLEU-4, ROUGE-L, METEOR, CIDEr, SPICE, and SPIDEr.

\vspace{5pt}
\noindent
\textbf{Training details:} All captions were tokenized using the word tokenizer of the natural language toolkit (NLTK) \cite{nltk}. All tokens in the development dataset were then counted, and words that appeared more than five times were appended in the word vocabulary. The vocabulary size was 2145, which includes BOS, EOS, PAD, and UNK tokens. The part-of-speech (POS)--tagging and lemmatization for keyword extraction were carried out using the POS--tagger and the WordNet Lemmatizer of the NLTK, respectively. Then, the most frequent $C=50$ lemmas were appended to the keyword vocabulary. The average number of keywords per caption was $2.23$, and we used $K=5$ because the number of keywords of 95\% of the training samples was less than five.

The encoder and decoder of TRACKE are composed of a stack of three identical layers, and each layer's multi-head attention/self-attention has four heads. All parameters in TRACKE were initialized using a random number from $\mathcal{N} (0, 0.02)$ \cite{init}. The number of hidden units was $D_f = 100$, and the initial and encoder/decoder's dropout probability were $0.5$ and $0.3$, respectively. We used the Adam optimizer \cite{adam} with $\beta_1 = 0.9$, $\beta_2 = 0.999$, and $\epsilon = 10^{-8}$ and varied the learning rate as the same formula of the original Transformer \cite{transformer}. TRACKE was trained for 300 epochs with a batch size of 100, and the best validation model was used as the final output.

\vspace{5pt}
\noindent
\textbf{Comparison methods:} TRACKE ($\mathtt{Ours}$) was compared with three other models:
\vspace{-3pt}
\begin{description}
 \setlength{\parskip}{0cm} 
 \setlength{\itemsep}{0.05cm} 
\item[$\mathtt{Baseline}$] The baseline model of the DCASE 2020 Challenge Task 6 \cite{ac1}.
\item[$\mathtt{LSTM}$] Long short-term memory (LSTM)--based seq2seq model \cite{seq2seq1,seq2seq2}. $\mathcal{E}$ was two-layer bidirectional-LSTM, and its outputs were aggregated by an attention layer. $\mathcal{D}$ was one-layer LSTM whose initial hidden state was the encoder output. The number of hidden units was 180.
\item[$\mathtt{Transformer}$] Transformer-based AAC. Its architecture is the same as TRACKE, except that the keyword-estimation branch was removed.
\end{description}
\vspace{-3pt}
To investigate the effect of the number of keywords $K$, we also evaluated TRACKE with $K=10$ ($\mathtt{Ours} (K=10)$), where $K=10$ was larger than the maximum number of keywords per audio clip in the training data. To confirm the upper-bound performance of TRACKE, we also compared it with two other models. One is $\mathtt{Oracle1}$; instead of $\hat{\bm{m}}$, the keywords in the meta-data of the Clotho dataset (i.e.\, Freesound tags) are passed to the decoder in both training/test stages, and the other is $\mathtt{Oracle2}$; instead of $\hat{\bm{m}}$, all 5 ground-truth captions of $\bm{m}$ is passed to the decoder in both training/test stages. $\mathtt{Oracle1}$ gives the oracle performance when the keywords are given manually, and $\mathtt{Oracle2}$ gives this when the estimation accuracy of the keywords is perfect.

\subsection{Results}

\begin{table*}[ttt]
\caption{Experimental results on Clotho dataset with DCASE2020 Challenge metrics}
\label{tab:example}
\centering
\begin{tabular}{ l c | ccccccccc }
\toprule
\textbf{Model} 	& \# of params.& \textbf{B-1}	& \textbf{B-2}	& \textbf{B-3}	& \textbf{B-4}	& \textbf{CIDEr}	& \textbf{METEOR}	& \textbf{ROUGE-L} & \textbf{SPICE} & \textbf{SPIDEr} \\	
\midrule
$\mathtt{Baseline}$		& 4.64M		& 38.9		& 13.6		& 5.5			& 1.5			& 7.4			& 8.4			& 26.2	& 3.3 	& 5.4 \\
$\mathtt{LSTM}$	 		& 1.12M		& 49.4		& 28.5		& 16.9		& 10.0		& 22.2		& 14.5		& 33.4	& 9.0	 	& 15.6 \\
$\mathtt{Transformer}$		& 1.11M		& 50.2		& 29.9		& 18.3		& 10.2		& 23.3		& 14.1		& 33.7	& 9.1 	& 16.2 \\
$\mathtt{Ours}(K=10)$ 			& 1.13M		& 49.9		& 29.7		& 18.4		& $\bm{10.8}$		& 23.0		& 14.5		& $\bm{34.5}$	& 9.1 	& 16.1 \\ 
$\mathtt{Ours}$	& 1.13M		& $\bm{52.1}$	& $\bm{30.9}$	& $\bm{18.8}$	& 10.7	& $\bm{25.8}$	& $\bm{14.9}$	& 34.2 		& $\bm{9.7}$	& $\bm{17.7}$ \\ 
\midrule 
$\mathtt{Oracle1}$ 	 & 1.11M		& 53.4		& 32.2		& 20.0		& 11.7		& 27.5		& 15.4		& 35.1	& 10.1 	& 18.8 \\
$\mathtt{Oracle2}$ 			& 1.11M		& 56.7		& 37.5		& 24.8		& 15.9		& 34.7		& 18.1		& 39.1	& 12.3 	& 23.5 \\ 
\bottomrule
\end{tabular}
\end{table*}

\begin{figure*}[t]
\centering
\includegraphics[width=170mm,clip]{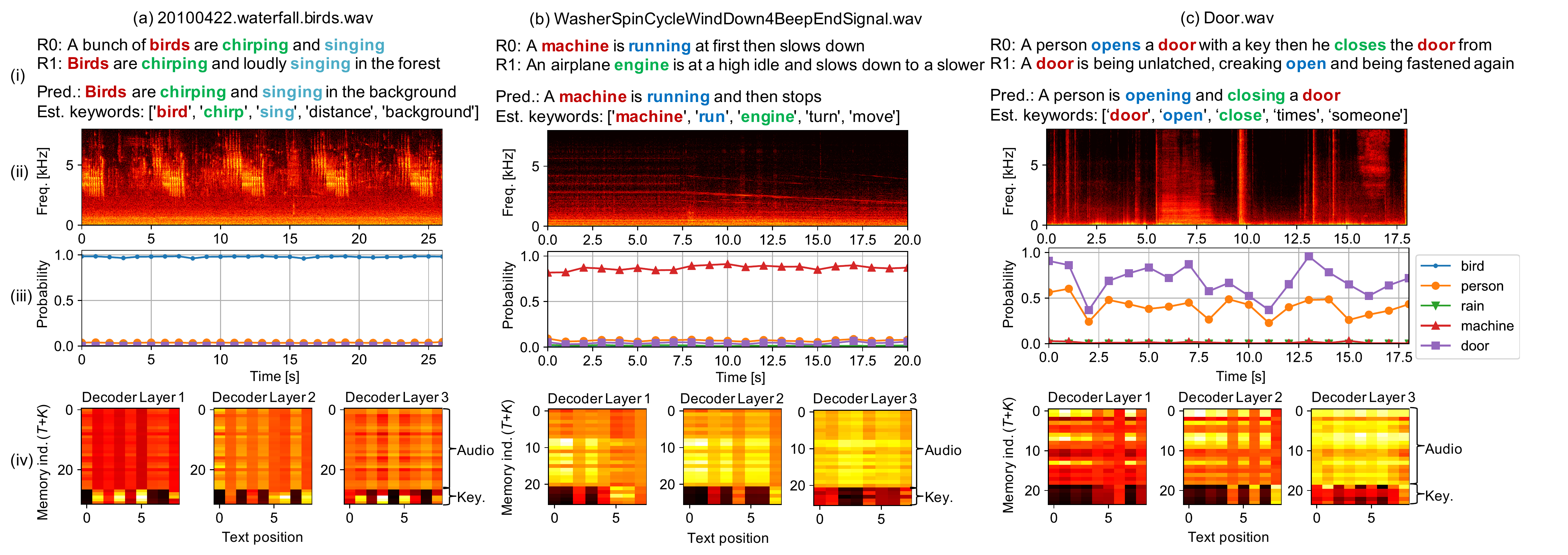} 
\vspace{-10pt}
\caption{Examples of TRACKE outputs. (i) Ground truth (R0 and R1) and estimated caption (Pred.) and keywords (Est. keywords), (ii) input spectrogram, (iii) keyword posterior of each time index $p( z_{c,t} | \bm{ \psi } )$, and (iv) attention matrices of decoder.}
 \label{fig:analysis}
 \vspace{-10pt}
\end{figure*}

Table \ref{tab:example} shows the evaluation results on the Clotho dataset. These results suggest the following:

(i) $\mathcal{M}$ works effectively for AAC. TRACKE ($\mathtt{Ours}$) achieved the highest score without given keywords. In addition, the BLEU-1 (ROUGE-L) score of $\mathtt{Ours}$ was 52.1 (34.2), while that of $\mathtt{Oracle1}$, which uses manually given keywords, was 53.4 (35.1). Thus, the score of $\mathtt{Ours}$ was 97.6\% (97.4\%) compared with $\mathtt{Oracle1}$, in spite the fact that $\mathtt{Ours}$ is a perfectly automated audio-captioning model.

(ii) If TRACKE can accurately estimate the keywords, performance might further improve. The oracle performance of $\mathtt{Oracle2}$ was significantly higher than that of $\mathtt{Ours}$. Since the keyword estimation accuracy of $\mathtt{Ours}$ was 48.1\%\footnote{The percentage of estimated keywords that were included in the ground-truth keywords.}, we need to improve this in future work.

(iii) If the estimated $K$ is too large, the use of the estimated keywords in text generation might be ineffective in reducing indeterminacy in word selection because the scores of $\mathtt{Transformer}$ and $\mathtt{Ours}(K=10)$ were almost the same. To further improve the performance of TRACKE, $K$ should also be estimated from the input.

(iv) Transformer might be effective for AAC because $\mathtt{Transformer}$ was slightly better than $\mathtt{LSTM}$. However, since the training of Transformer requires a large-scale dataset, to affirm the effectiveness of Transformer, we need to evaluate Transformer by developing more large-scale datasets for AAC\footnote{The number of training sentence pairs in natural language processing datasets, such as WMT 2014 English-French dataset, for machine translation is 36 million.}.

(v) The use of pre-trained models is effective because there were large performance gaps between $\mathtt{Baseline}$ and the others, and the major difference was the use of pre-trained models such as VGGish \cite{vggish} and fastText \cite{fastText}.

Figure \ref{fig:analysis} shows examples of TRACKE outputs. These results suggest that indeterminacy words were determined while referring to the estimated keywords, for example, (b) \{machine, airplane\} and (c) \{close, fasten\}. In addition, the posterior probabilities of keywords imply the implicit co-occurrence relationships, rather than just classifying acoustic events/scenes. In (c), the posterior probability of ``person'' increased even though human sounds, such as speech, were not included in the input audio. This might be the result of exploiting the co-occurrence relationship that opening and closing a door is usually done by humans.

\section{Conclusions}
\label{sec:cncl}
We proposed a Transformer-based audio captioning model with keyword estimation called \textit{TRACKE}, which simultaneously solves the word-selection indeterminacy problem of the main task of ACC while executing the AED/ASC- sub-task (i.e.\, keyword estimation). TRACKE estimates the keywords of the target caption from input audio, and its decoder generates a caption while referring to the estimated keywords. The keyword-estimation branch was trained by adopting a weakly supervised polyphonic AED strategy \cite{wpoly}, and the ground-truth keywords were extracted from the ground-truth caption via a heuristic rule. The experimental results indicate the effectiveness of TRACKE for AAC.

Future work includes improving keyword estimation while adopting keyword-guided generation strategies in natural language processing \cite{keyword01,keyword02,keyword03,keyword05,keyword06} and image captioning \cite{im_cap1,im_cap2,im_cap3,im_cap4}.

\newpage
\bibliographystyle{IEEEtran}

\end{document}